\documentclass[conference]{IEEEtran}
\IEEEoverridecommandlockouts

\usepackage{cite}
\usepackage{amsmath,amssymb,amsfonts}
\usepackage{algorithmic}
\usepackage{graphicx}
\usepackage{textcomp}
\usepackage{xcolor}
\usepackage{booktabs}
\usepackage{tikz}
\usetikzlibrary{shapes.geometric,arrows.meta,positioning}
\usepackage{array}
\usepackage{url}
\usepackage[hidelinks]{hyperref}
\usepackage{orcidlink}

\def\BibTeX{{\rm B\kern-.05em{\sc i\kern-.025em b}\kern-.08em T\kern-.1667em\lower.7ex\hbox{E}\kern-.125emX}}

\begin{document}

\title{HIPAA-Compliant AI Deployment Patterns in\\Clinical Settings: Privacy-Preserving Techniques\\and Governance Controls}

\author{
\IEEEauthorblockN{Vinod Dhiman~\orcidlink{0009-0003-8358-9382}}
\IEEEauthorblockA{\textit{Technical Program Manager} \\
Arlington, VA, USA \\
vinod.dhiman@icloud.com}
\thanks{Corresponding author: Vinod Dhiman (e-mail: vinod.dhiman@icloud.com).
ORCID: 0009-0003-8358-9382.}
}

\maketitle

\begin{abstract}
Artificial intelligence (AI) is moving from research prototypes into
clinical workflows, yet the deployment of AI systems that process
protected health information (PHI) remains constrained by the U.S.
Health Insurance Portability and Accountability Act (HIPAA) and by the
absence of shared engineering guidance for satisfying it. Existing work
tends to treat privacy-preserving machine learning and regulatory
governance as separate concerns, leaving practitioners without a
concrete mapping from architectural decisions to compliance obligations.
This paper contributes a catalog of five reusable deployment patterns
for clinical AI that jointly address the HIPAA Privacy and Security
Rules: a de-identified analytics zone, cross-entity federated learning,
confidential inference within trusted execution environments, a
PHI-minimizing retrieval-augmented clinical assistant, and a synthetic
data sandbox for development. Each pattern is specified in terms of
context, governing forces, structure, and residual risk, and is mapped
to the administrative, physical, and technical safeguards of the HIPAA
Security Rule as well as to the NIST AI Risk Management Framework and
ISO/IEC~42001. We further present a threat model for clinical AI and a
decision procedure for selecting patterns based on data sensitivity,
trust boundaries, and latency requirements. The goal is to give
security engineers, compliance owners, and clinical informaticists a
common vocabulary that turns abstract regulatory duties into verifiable
architectural controls.
\end{abstract}

\begin{IEEEkeywords}
HIPAA, protected health information, privacy-preserving machine
learning, differential privacy, federated learning, confidential
computing, AI governance, NIST AI RMF, ISO/IEC 42001, clinical
informatics
\end{IEEEkeywords}

\section{Introduction}
The clinical adoption of artificial intelligence has accelerated across
diagnostics, clinical documentation, risk stratification, and
operational optimization~\cite{topol-medicine}. Large language models now
draft clinical notes, summarize charts, and support prior-authorization
workflows, while predictive models forecast deterioration, readmission,
and sepsis.
Nearly all of these systems consume protected health information (PHI):
identifiers, diagnoses, medications, images, and free-text narratives
that are among the most sensitive categories of personal data.

In the United States, the processing of PHI by covered entities and
their business associates is governed by HIPAA, whose Privacy Rule
constrains permissible uses and disclosures and whose Security Rule
mandates administrative, physical, and technical safeguards for
electronic PHI (ePHI)~\cite{hipaa-privacy,hipaa-security}. HIPAA was
written long before modern machine learning and does not speak directly
to model training, inference serving, vector databases, or prompt
logging. As a result, teams that build clinical AI must translate
technology-neutral legal requirements into concrete architectural
decisions, often without a shared reference. The consequences of getting
this translation wrong are severe: unauthorized disclosure, enforcement
penalties, breach-notification obligations, and erosion of patient
trust.

Two research communities offer partial answers. The
privacy-preserving machine learning (PPML) community has produced a rich
toolkit---differential privacy, federated learning, homomorphic
encryption, secure multiparty computation, and synthetic data
generation~\cite{dwork-dp,mcmahan-fl,acar-he}. Separately, the AI
governance community has produced management frameworks such as the NIST
AI Risk Management Framework (AI RMF)~\cite{nist-airmf} and ISO/IEC
42001~\cite{iso42001}, which specify how organizations should govern AI
risk. What is missing is the connective tissue: a set of concrete,
reusable deployment patterns that show how specific PPML techniques,
composed into an architecture, satisfy specific HIPAA safeguards and
governance controls.

This paper aims to supply that connective tissue. Our contributions are:

\begin{itemize}
\item A concise threat model for AI systems that process PHI, organized
around the trust boundaries that matter for HIPAA compliance
(Section~\ref{sec:threat}).
\item A structured review of privacy-preserving techniques oriented
toward their compliance value rather than their algorithmic novelty
(Section~\ref{sec:techniques}).
\item A catalog of five deployment patterns for clinical AI, each
specified with context, forces, structure, and residual risk
(Section~\ref{sec:patterns}).
\item A mapping from each pattern to the HIPAA Security Rule safeguards,
the NIST AI RMF functions, and ISO/IEC 42001 controls, together with a
selection procedure (Sections~\ref{sec:governance}
and~\ref{sec:discussion}).
\end{itemize}

We emphasize that patterns are not a substitute for legal review or a
formal risk analysis; they are engineering scaffolding that makes such
reviews faster, more consistent, and more auditable.

\section{Background}
\label{sec:background}

\subsection{HIPAA in Brief}
HIPAA's Privacy Rule establishes when PHI may be used or disclosed and
enshrines the \emph{minimum necessary} principle: uses and disclosures
should be limited to the least amount of PHI needed to accomplish the
purpose. The Security Rule applies specifically to ePHI and organizes
its requirements into three families of safeguards. Administrative
safeguards include security management, workforce training, and
contingency planning. Physical safeguards address facility access and
device controls. Technical safeguards include access control, audit
controls, integrity protection, and transmission
security~\cite{hipaa-security}. Business associate agreements (BAAs)
extend these obligations contractually to vendors---including cloud
providers and AI platforms---that handle ePHI on behalf of covered
entities.

\subsection{De-identification Standards}
HIPAA recognizes two routes to de-identified data, which falls outside
the scope of the Privacy Rule. The \emph{Safe Harbor} method requires
the removal of eighteen enumerated identifier categories. The
\emph{Expert Determination} method requires a qualified statistician to
certify that the re-identification risk is very small~\cite{hhs-deid}.
Both routes are central to clinical AI because de-identified data can be
used for model development with far fewer restrictions---though free-text
notes and imaging make robust de-identification technically demanding,
often requiring neural named-entity recognition to detect identifiers
embedded in narrative text~\cite{dernoncourt-deid}, and
re-identification research shows that naive removal of explicit
identifiers is often insufficient~\cite{sweeney-reident}.

\subsection{Why Clinical AI Strains the Model}
Machine learning introduces data flows that traditional HIPAA controls
did not anticipate. Training aggregates PHI at scale; models can
memorize and later leak training examples~\cite{carlini-extract};
retrieval-augmented generation copies PHI into vector stores and prompt
contexts; and inference logs, telemetry, and evaluation datasets create
new copies of ePHI in places that are easy to overlook. Each of these
flows is a candidate for a safeguard, and each is addressed by one or
more of the patterns below.

\section{Threat Model for Clinical AI}
\label{sec:threat}
We frame threats around the trust boundaries an auditor would examine.
The assets are ePHI at rest (training corpora, vector indices, logs),
ePHI in transit (API calls, inter-service traffic), and the model
itself as a potential carrier of memorized PHI. We consider four
adversary classes:

\begin{itemize}
\item \textbf{External attacker} who compromises an exposed endpoint,
storage bucket, or credential and seeks bulk PHI exfiltration.
\item \textbf{Malicious or negligent insider} with legitimate platform
access who exceeds the minimum-necessary boundary.
\item \textbf{Curious cloud or platform operator} outside the covered
entity's administrative control, motivating confidentiality guarantees
that hold even against the infrastructure host.
\item \textbf{Inference adversary} who interacts only with the deployed
model and attempts membership-inference, attribute-inference, model
inversion, or training-data-extraction
attacks~\cite{shokri-mia,carlini-extract,fredrikson-inversion}.
\end{itemize}

Table~\ref{tab:threats} summarizes representative threats and the
techniques that mitigate them. This mapping motivates the pattern
catalog: no single technique addresses all adversaries, so patterns
compose techniques to close the gaps relevant to a given deployment.

\begin{table}[t]
\caption{Representative threats and mitigating techniques}
\label{tab:threats}
\centering
\renewcommand{\arraystretch}{1.25}
\begin{tabular}{@{}p{0.30\columnwidth} p{0.58\columnwidth}@{}}
\toprule
\textbf{Threat} & \textbf{Primary mitigations} \\
\midrule
Bulk exfiltration at rest & Encryption, de-identification, tokenization, access control, audit \\
PHI in inference logs & Log minimization, redaction, short retention \\
Cross-site data pooling & Federated learning, secure aggregation \\
Untrusted host access & Confidential computing, homomorphic encryption \\
Training-data leakage via model & Differential privacy, output filtering \\
Membership inference & DP training, regularization, query limits \\
\bottomrule
\end{tabular}
\end{table}

\section{Privacy-Preserving Techniques}
\label{sec:techniques}
We briefly review the building blocks, emphasizing their compliance role.

\textbf{De-identification and tokenization.} Removing or surrogating
identifiers is the first line of defense and the only technique that can
move data out of HIPAA scope entirely. In practice, clinical text
requires named-entity recognition to detect identifiers embedded in
narratives, and imaging requires pixel- and metadata-level scrubbing.
Tokenization replaces identifiers with reversible surrogates held in a
separately controlled vault, preserving referential integrity for
operations while limiting exposure.

\textbf{Differential privacy (DP).} DP provides a formal, quantifiable
bound on how much any single patient's data can influence an
output~\cite{dwork-dp}. Applied during training via DP-SGD~\cite{abadi-dpsgd},
it limits memorization and defends against membership-inference and
extraction attacks, at a measurable cost to model utility governed by
the privacy budget $\varepsilon$.

\textbf{Federated learning (FL).} FL trains a shared model across
institutions without centralizing raw PHI; only model updates
leave each site~\cite{mcmahan-fl,kairouz-fl}, an approach demonstrated
for multi-institutional clinical modeling without pooling patient
data~\cite{sheller-fl}. Combined with secure
aggregation~\cite{bonawitz-secagg}, the coordinating server never sees
individual updates in the clear, which addresses cross-site pooling
concerns and can reduce the data-sharing footprint that BAAs must cover.

\textbf{Homomorphic encryption (HE) and secure multiparty computation
(SMPC).} HE permits computation directly on
ciphertext~\cite{acar-he,gentry-fhe}, enabling inference over encrypted
inputs~\cite{gilad-cryptonets}, and SMPC lets mutually distrusting
parties jointly compute a function without revealing their
inputs~\cite{yao-smpc,mohassel-secureml}. Both offer strong
confidentiality against an untrusted host but impose substantial
performance overhead, so they are best reserved for narrow,
high-sensitivity computations.

\textbf{Confidential computing.} Hardware trusted execution environments
(TEEs) isolate code and data in encrypted memory, allowing PHI to be
processed with attestable protection even from the platform
operator~\cite{costan-sgx}, and can be combined with cryptographic
verification to protect neural-network execution in
hardware~\cite{tramer-slalom}. TEEs offer a more favorable performance
profile than HE for full-model inference, at the cost of trusting the
hardware root of trust.

\textbf{Synthetic data.} Generative models can produce artificial
records that preserve statistical structure while, ideally, containing
no real patient's data. Synthetic data is valuable for development,
testing, and demonstration, but privacy guarantees are only as strong as
the generation process; DP-trained generators are preferred, and
membership-inference evaluation should gate release.

\section{Deployment Patterns}
\label{sec:patterns}
We now present the core contribution: five deployment patterns. Each is
described by \emph{context} (when it applies), \emph{forces} (the
competing pressures), \emph{structure} (the architecture), and
\emph{residual risk} (what remains and must be governed).

\subsection{Pattern 1: De-identified Analytics Zone}
\textbf{Context.} Population-level analytics, cohort discovery, and model
development that do not require patient-level identifiers.

\textbf{Forces.} Teams want broad, low-friction access to data for
experimentation, but identifiable PHI carries the full weight of the
Privacy Rule and the minimum-necessary constraint.

\textbf{Structure.} A one-way pipeline ingests source ePHI into a
controlled landing zone, applies Safe Harbor or Expert Determination
de-identification (with NER-based scrubbing for free text), and lands the
result in a separate analytics environment with its own access
controls. A tokenization vault, isolated under stricter controls,
retains any re-linkage capability needed for authorized operational use.

\textbf{Residual risk.} Re-identification risk from quasi-identifiers and
residual identifiers in unstructured text; mitigated by Expert
Determination review, quasi-identifier generalization, and periodic
re-identification testing.

\subsection{Pattern 2: Cross-Entity Federated Learning}
\textbf{Context.} Multiple covered entities want a jointly trained model
(e.g., a rare-disease classifier) but cannot or will not centralize
their PHI.

\textbf{Forces.} Model quality improves with more diverse data, yet
pooling data multiplies exposure and complicates the BAA landscape.

\textbf{Structure.} Each site trains locally on its own ePHI; a
coordinator aggregates model updates using secure aggregation so that no
individual site's update is visible in the clear. DP noise can be added
to updates to bound leakage through the shared model. Only model
parameters cross institutional boundaries.

\textbf{Residual risk.} Update-based leakage and reconstruction attacks;
mitigated by secure aggregation, DP, and gradient clipping. Governance
must still define accountability for the shared model's outputs.

\subsection{Pattern 3: Confidential Inference Enclave}
\textbf{Context.} Real-time inference on identifiable PHI where the
processing environment is not fully within the covered entity's
administrative control, such as a shared or third-party platform.

\textbf{Forces.} Latency requirements rule out heavyweight cryptography,
but the untrusted-host threat demands protection beyond transport
encryption.

\textbf{Structure.} The model runs inside a hardware TEE. Clients verify
a remote attestation before transmitting ePHI over an encrypted channel
terminated inside the enclave. Keys are released only to attested
enclaves. Plaintext PHI exists only within encrypted enclave memory and
is never persisted outside it.

\textbf{Residual risk.} Trust in the hardware vendor and side-channel
exposure; mitigated by attestation policy, patching, and minimizing
enclave-resident data lifetime.

\subsection{Pattern 4: PHI-Minimizing Clinical Assistant}
\textbf{Context.} Retrieval-augmented generation (RAG)~\cite{lewis-rag}
assistants that answer clinician questions over patient records or
knowledge bases.

\textbf{Forces.} RAG improves accuracy by injecting patient context into
prompts, but doing so copies PHI into vector stores, prompt logs, and
possibly third-party model providers.

\textbf{Structure.} A minimization layer sits between the record system
and the model. It enforces per-request scoping to the minimum necessary
records, de-identifies or tokenizes context where the task allows,
applies output filters to catch inadvertent identifier disclosure, and
disables or redacts prompt/response logging for PHI. The vector store
holding embeddings of clinical text is treated as ePHI and secured
accordingly, with a BAA covering any external model endpoint.

\textbf{Residual risk.} Prompt-injection-driven over-disclosure and
embedding-inversion leakage~\cite{owasp-llm}; mitigated by strict
retrieval scoping, input/output guardrails, and treating embeddings as
sensitive.

\subsection{Pattern 5: Synthetic Data Sandbox}
\textbf{Context.} Development, testing, demonstrations, and vendor
evaluations that should never touch real PHI.

\textbf{Forces.} Developers need realistic data to build and debug, but
every additional copy of PHI in lower environments expands the attack
surface.

\textbf{Structure.} A DP-trained generator, fitted in a controlled
environment, produces synthetic records that populate development and
test environments. A privacy gate runs membership-inference and
nearest-neighbor tests before any synthetic dataset is promoted for
wider use. Lower environments are provisioned with synthetic data by
default and are contractually and technically barred from real ePHI.

\textbf{Residual risk.} Utility loss and residual leakage from
overfit generators; mitigated by DP generation, privacy gating, and
periodic re-validation.

\section{Mapping to Governance Controls}
\label{sec:governance}
Patterns become auditable only when tied to named controls.
Table~\ref{tab:mapping} maps each pattern to the HIPAA Security Rule
safeguard families and to the primary NIST AI RMF function it advances.
The AI RMF organizes activity into \textsc{Govern}, \textsc{Map},
\textsc{Measure}, and \textsc{Manage}~\cite{nist-airmf}; ISO/IEC 42001
provides the surrounding management-system structure and Annex~A
controls for the AI lifecycle~\cite{iso42001}.

\begin{table}[t]
\caption{Patterns mapped to HIPAA safeguards and NIST AI RMF}
\label{tab:mapping}
\centering
\renewcommand{\arraystretch}{1.25}
\footnotesize
\begin{tabular}{@{}p{0.30\columnwidth} p{0.30\columnwidth} p{0.24\columnwidth}@{}}
\toprule
\textbf{Pattern} & \textbf{HIPAA safeguards} & \textbf{AI RMF} \\
\midrule
De-identified zone & Access control; minimum necessary & Map, Manage \\
Federated learning & Transmission security; access control & Measure, Manage \\
Confidential enclave & Access control; transmission security; integrity & Manage \\
Clinical assistant & Access, audit, minimum necessary & Measure, Manage \\
Synthetic sandbox & Access control; security management & Govern, Map \\
\bottomrule
\end{tabular}
\end{table}

Beyond the technical safeguards, three administrative controls cut
across all patterns. First, a documented \emph{risk analysis} that
enumerates ePHI flows for each AI system, refreshed when models or data
flows change, following established guidance for the HIPAA Security
Rule~\cite{nist-sp80066}. Second, \emph{business associate agreements} that
explicitly cover model hosting, training, logging, and evaluation
activities, not merely storage. Third, \emph{audit and monitoring}
sufficient to reconstruct who accessed which PHI through the AI system,
including prompts and retrieved context, which are frequently omitted
from logging designs. ISO/IEC 42001 adds lifecycle controls---impact
assessment, data governance, and post-deployment monitoring---that align
the AI system with the organization's stated objectives and risk
appetite.

\section{Discussion: Selecting a Pattern}
\label{sec:discussion}
\begin{figure}[t]
\centering
\footnotesize
\begin{tikzpicture}[
  node distance=6.5mm and 5mm,
  every node/.style={font=\footnotesize},
  dec/.style={diamond, aspect=2, draw, align=center, inner sep=1pt,
              text width=20mm},
  pat/.style={rectangle, rounded corners, draw, align=center,
              inner sep=3pt, text width=22mm},
  arr/.style={-{Latex[length=2mm]}}
]
\node[dec] (id) {Needs patient identifiers?};
\node[pat, below left=of id] (deid) {De-identified zone / Synthetic sandbox};
\node[dec, below right=of id] (bound) {Crosses a trust boundary?};
\node[pat, below left=8mm and -2mm of bound] (fl) {Federated learning};
\node[dec, below right=of bound] (host) {Untrusted host?};
\node[pat, below left=8mm and -4mm of host] (enc) {Confidential enclave};
\node[pat, below=of host] (asst) {PHI-minimizing assistant};

\draw[arr] (id) -- node[above left]{no} (deid);
\draw[arr] (id) -- node[above right]{yes} (bound);
\draw[arr] (bound) -- node[above left]{cross-site} (fl);
\draw[arr] (bound) -- node[above right]{single site} (host);
\draw[arr] (host) -- node[above left]{yes} (enc);
\draw[arr] (host) -- node[right]{no} (asst);
\end{tikzpicture}
\caption{Decision procedure for selecting a deployment pattern based on
identifier need, trust boundaries, and host control.}
\label{fig:decision}
\end{figure}
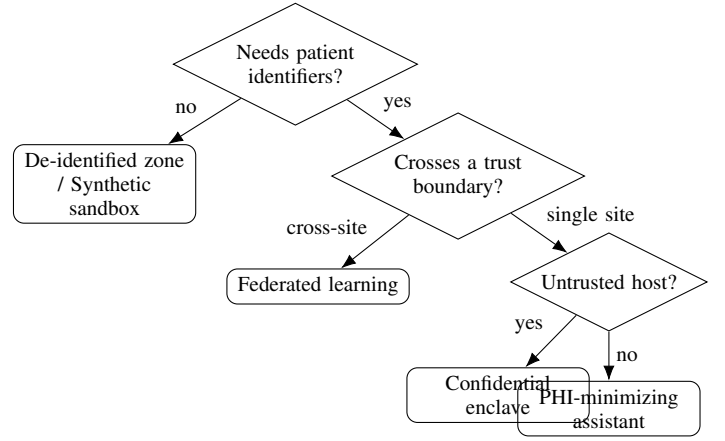

The patterns are complementary and frequently combined. A realistic
clinical AI program might use the synthetic sandbox for development, the
de-identified zone for model training, and a confidential enclave or
PHI-minimizing assistant for production inference. To guide selection,
we suggest a short decision procedure driven by three questions, shown in
Fig.~\ref{fig:decision}.

\emph{Does the task require patient-level identifiers?} If not, prefer
the de-identified zone or synthetic sandbox, which remove or reduce
HIPAA scope. \emph{Is raw data crossing an institutional or trust
boundary?} If institutions must collaborate without pooling data, choose
federated learning; if a single institution uses an untrusted host,
choose the confidential enclave. \emph{Is the workload interactive over
identifiable records?} If so, the PHI-minimizing assistant pattern
supplies the minimization, guardrail, and logging discipline that
generative workflows demand.

Two cross-cutting principles govern all cases. \emph{Data minimization}
should be applied at every layer, because the cheapest PHI to protect is
the PHI that was never copied. \emph{Defense in depth} should be
assumed, because no single technique resists every adversary in
Table~\ref{tab:threats}; patterns earn their value precisely by
composing techniques so that the failure of one control does not expose
PHI.

A practical tension deserves acknowledgment. Stronger privacy techniques
frequently reduce model utility or increase cost: DP trades accuracy for
a bounded privacy budget, HE and SMPC trade latency for
host-independence, and federated learning trades convergence speed and
engineering simplicity for reduced data movement. The right operating
point is a governance decision, not purely a technical one, and should
be recorded in the risk analysis so that the accepted residual risk is
explicit and revisitable.

\subsection{Worked Example}
Consider a health system that wants to deploy an ambient documentation
assistant that drafts clinical notes from patient encounters, and
separately wants to build a multi-site early-warning model for clinical
deterioration. A single program can compose four of the five patterns.

During development, engineers work exclusively against the
\emph{synthetic sandbox} (Pattern~5): a DP-trained generator populates
the test environment with realistic but artificial encounters, and a
privacy gate blocks promotion of any synthetic dataset that fails
membership-inference screening. No real ePHI reaches lower environments,
collapsing a large swath of the attack surface that would otherwise
require safeguards.

To build the deterioration model, the system collaborates with two peer
institutions using \emph{cross-entity federated learning} (Pattern~2).
Each site trains locally; secure aggregation hides individual updates
from the coordinator, and DP noise bounds what the shared model can leak.
The BAA landscape shrinks because raw records never leave any site. For
retrospective analytics on the resulting cohorts, identifiers are
unnecessary, so the team routes data through the \emph{de-identified
analytics zone} (Pattern~1) rather than granting analysts access to
identifiable PHI.

In production, the ambient assistant is an interactive workload over
identifiable records, so it adopts the \emph{PHI-minimizing clinical
assistant} pattern (Pattern~4): per-request scoping to the minimum
necessary records, output guardrails against inadvertent identifier
disclosure, embeddings treated as ePHI, and PHI-aware logging controls.
Because the drafting model is hosted on a shared platform outside the
system's direct administrative control, inference runs inside a
\emph{confidential enclave} (Pattern~3), with attestation gating key
release. The result is a single program in which each data flow is
matched to a pattern, each pattern to a set of safeguards, and each
accepted residual risk to a line in the risk analysis.

\subsection{Breach and Incident Considerations}
Pattern selection also shapes breach exposure. Data that has been
properly de-identified or replaced with vetted synthetic records is not
PHI, so incidents involving those stores generally fall outside
breach-notification obligations---one of the strongest practical
arguments for pushing minimization as far upstream as possible.
Conversely, the newer artifacts introduced by AI systems---vector
indices, prompt and completion logs, and evaluation datasets---are easy
to omit from an incident-response plan precisely because they are not
recognized as ePHI stores. A defensible clinical AI program inventories
these artifacts explicitly, assigns them the same containment and
notification procedures as any other ePHI repository, and rehearses
their inclusion in tabletop exercises.

\section{Operationalizing the Patterns}
\label{sec:ops}
A pattern is auditable only if it produces evidence. In our experience,
the gap between an intended control and a defensible one is the artifact
that demonstrates the control operated as designed. Table~\ref{tab:evidence}
lists representative evidence artifacts an auditor or internal reviewer
would expect for each pattern. Framing the artifacts up front, before the
system ships, converts compliance from a retrospective scramble into a
byproduct of normal engineering.

\begin{table}[t]
\caption{Representative audit evidence per pattern}
\label{tab:evidence}
\centering
\renewcommand{\arraystretch}{1.3}
\footnotesize
\begin{tabular}{@{}p{0.28\columnwidth} p{0.62\columnwidth}@{}}
\toprule
\textbf{Pattern} & \textbf{Evidence artifacts} \\
\midrule
De-identified zone &
Expert Determination report or Safe Harbor checklist; NER scrubbing
metrics; re-identification test results; vault access logs \\
Federated learning &
Secure-aggregation configuration; per-round DP budget accounting;
proof that raw data never left each site \\
Confidential enclave &
Attestation policy and verification logs; key-release records; enclave
memory-lifetime configuration \\
Clinical assistant &
Minimum-necessary scoping rules; guardrail test suite; logging/redaction
configuration; embedding-store classification \\
Synthetic sandbox &
Generator DP parameters; membership-inference gate results; environment
policy barring real ePHI \\
\bottomrule
\end{tabular}
\end{table}

These artifacts also connect the technical layer back to the management
system. Under ISO/IEC~42001, they become inputs to the AI impact
assessment and to periodic management review; under the NIST AI RMF, they
substantiate the \textsc{Measure} and \textsc{Manage} functions; and under
HIPAA, they populate the evidence base a covered entity would present in
the event of an Office for Civil Rights inquiry. Designing the evidence
alongside the architecture is what makes a pattern defensible rather than
merely plausible.

\section{Related Work}
\label{sec:related}
Privacy-preserving machine learning has a mature literature spanning
differential privacy~\cite{dwork-dp,abadi-dpsgd}, federated
learning~\cite{mcmahan-fl,kairouz-fl,bonawitz-secagg}, and cryptographic
computation~\cite{acar-he,gentry-fhe}, and a substantial body of work
documents attacks that motivate these defenses, including membership
inference~\cite{shokri-mia}, model inversion~\cite{fredrikson-inversion},
and training-data extraction~\cite{carlini-extract}. In the healthcare
domain specifically, federated and privacy-preserving learning have been
applied to clinical and medical-imaging workloads while preserving
institutional data boundaries~\cite{rieke-fl-health,kaissis-medimg,sheller-fl},
and prior work examines de-identification of clinical text and the limits
of anonymization~\cite{sweeney-reident,hhs-deid}. On the governance side,
the NIST AI RMF~\cite{nist-airmf} and ISO/IEC 42001~\cite{iso42001}
provide organizational frameworks, cross-framework taxonomies aim to
unify fragmented AI governance controls~\cite{uagt}, and regulatory
guidance addresses HIPAA obligations~\cite{hipaa-privacy,hipaa-security}.
Our contribution is orthogonal and integrative: rather than advancing a
single technique or framework, we package techniques into deployment
patterns and bind them explicitly to HIPAA safeguards and governance
functions, giving practitioners a reusable bridge between the two
literatures.

\section{Limitations and Future Work}
\label{sec:limitations}
This work is conceptual and does not yet include an empirical evaluation
of the patterns in a production clinical environment; validating utility,
latency, and cost trade-offs across representative workloads is important
future work. The patterns are framed around U.S. HIPAA requirements and
would need adaptation for other regimes such as the EU AI
Act~\cite{eu-aiact} and the GDPR~\cite{gdpr}, though the underlying
techniques transfer. Finally, the governance
mapping is intended as a starting point for a formal risk analysis, not
a certification; future work includes a machine-readable control
mapping and reference implementations that emit audit evidence
automatically.

\section{Conclusion}
\label{sec:conclusion}
Deploying AI on protected health information forces a translation from
technology-neutral regulation into concrete architecture. We have argued
that this translation is best captured as reusable deployment patterns
that compose privacy-preserving techniques and bind them to named
governance controls. The five patterns presented here---de-identified
analytics, federated learning, confidential inference, PHI-minimizing
assistants, and synthetic sandboxes---cover the dominant clinical AI
workloads and, taken together with the threat model and control mapping,
give security and compliance practitioners a shared vocabulary for
building clinical AI that is defensible by design rather than compliant
by afterthought.


\end{document}